\documentclass{amsart}
\date{June 6, 2003}
\usepackage{epsfig}
\newtheorem{theorem}{Theorem}

\newtheorem{remark}{Remark}

\def\ii{\'\i}

\title[Variational calculation of the period of nonlinear oscillators
]{Variational calculation of the period of nonlinear oscillators
 }
\begin{document}

\author[R.D.~Benguria]{Rafael Benguria}
\address{Pontificia Universidad Cat\'olica de Chile\\
  Departamento de F\ii sica\\
  Casilla 306\\
  Santiago 22\\
  Chile} \email{rbenguri@fis.puc.cl}

\author[M.C.~Depassier]{M. Cristina Depassier}
\address{Pontificia Universidad Cat\'olica de Chile\\
  Departamento de F\ii sica\\
  Casilla 306\\
  Santiago 22\\
  Chile} \email{mcdepass@puc.cl}

\dedicatory{{\large Dedicated to Elliott H. Lieb on his 70th birthday}}

\thanks{This work is partially supported by FONDECYT
  (Chile), projects 102--0844 and 102--0851.}

\begin{abstract}
The problem of calculating the period of second order nonlinear autonomous oscillators is formulated as an eigenvalue problem. We show that the period can be obtained from two integral variational principles dual to each other. Upper and lower bounds on the period can be obtained to any desired degree of accuracy. The results are illustrated by an application to the Duffing equation.

\end{abstract}

\maketitle

\section{Introduction}

The motion of a particle in one dimension under the action of a conservative force is described by the simple equation $\ddot u =-f(u)$, where the force $-f$ depends on the amplitude of the motion $u$ and one or more parameters. For such conservative systems the total energy $E = \dot u^2/2 + V(u)$, is conserved. While a complete description of the motion can be obtained from the analysis of phase space, an explicit analytic expression for the period of oscillations around equilibria is not always available. For linear problems the period does not depend on the amplitude of the motion, but when nonlinearities are present, even though we know the expression for the period, namely
\begin{equation}
T = 2 \int_{u_0}^{u_1} \frac{ d u}{\sqrt{2(E-V(u))}},
\label{eq:1.1}
\end{equation}
where $u_0$ and $u_1$ are turning points, the integral cannot be evaluated analytically except for special nonlinear potentials $V(u)$. The usual procedure in such cases is to consider the nonlinearities as small perturbations to the linear problem and to obtain a series expansion for the period of oscillation. To obtain a close estimate of the period at large amplitude, the perturbation expansion must be carried to high orders. Often the series may be divergent. Recently a method to obtain a rapidly convergent series has been developed \cite{PeKlSc03}, 
based on Kleinert's variational perturbation theory (see Chapter 5 of \cite{Kl03}).

In previous work we studied a nonlinear eigenvalue problem, of which the above problem constitutes a particular case, and showed that the eigenvalue (the period in the present case) can be obtained from an integral variational principle.  From it, we obtained arbitrarily accurate upper bounds on the period, valid at all amplitudes, without resource to perturbation theory. The purpose of the present work is to show the existence of a dual variational principle from which lower bounds are obtained, thus providing a complete variational characterization of the period.

We consider systems with an equilibrium point $u=0$, and will study the period of oscillations around this equilibrium. Moreover, for the sake of simplicity,  we assume that the force is an odd function of $u$. The results can be generalized in a simple way to a general force term. If the force is odd, the period can be evaluated by considering the motion through a quarter of a period. Chosing the quarter period in the quadrant $(\dot u <0, u>0)$ of phase space, the period may be calculated solving
$\ddot u =-f(u),$ with $u(0) = u_m$, $\dot u(0) =0$ and  $u(T/4) =0$.

Introducing the new variable $\tau = 4t/T$, the problem reduces to finding the eigenvalue $\lambda$ of
\begin{equation}
- u'' = \lambda f(u), \qquad \qquad u(0) = u_m, \qquad u'(0) = 0, \qquad u(1) =0.
\label{problem}
\end{equation}
The period is given by $T = 4\sqrt{\lambda}$. Here, primes denote derivatives with respect to $\tau$. Notice that the eigenvalue $\lambda$ depends on the amplitude $u_m$.

In previous work we proved that the eigenvalue $\lambda$ is characterized by the variational principle
\begin{equation}
\lambda [u_m] = \max_g \frac{1}{2} \frac{ \left( \int_0^{u_m} {g'}^{1/3}(u) d u
\right)^3}{\int_0^{u_m} f(u) g(u) d u},
\label{var1}
\end{equation}
 where the maximum is taken over all positive functions $g(u)$ such that $g(0) =0$, $g'(u) >0$. The maximum is achieved for $g = \hat g$ which satisfies
$$
\hat g'(u) = \frac{1}{(E-V(u))^{3/2}}.
$$
An application of this result to the Duffing equation yielded a simple approximate formula as a close upper bound for the frequency $\omega = 2\pi/T$. In the following we construct the dual of (\ref{var1}). The two variational principles give a complete characterization of the eigenvalue problem. In Section 2 we give the complete variational characterization of the period. In Section 3 we apply the variational principle and its dual to estimate the period of the Duffing oscillator.

\bigskip

\section{A variational principle and its dual}

Many authors have considered the following two point boundary value problem
\begin{equation}
-y''= \lambda f(y) \qquad \mbox{in $(0,1)$}
\label{eq:2.1}
\end{equation}
with 
\begin{equation}
y'(0)=0, \qquad y(1)=0, \qquad \mbox{and} \qquad  y(0) \equiv y_m.
\label{eq:2.2}
\end{equation}
Here $f(y)$ is positive and continuous but not necessarily $0$ when $y=0$. For example, this type of nonlinear two point boundary value problem arises in the study of heat generation and stability of temperature distribution of conducting plates \cite{Jo65}. Under the assumption that $f$ is positive and continuous, Laetsch \cite{La70} proved the existence and regularity of nontrivial solutions to (\ref{eq:2.1}),(\ref{eq:2.2}). In fact, the positive solutions to 
(\ref{eq:2.1}),(\ref{eq:2.2}) are decreasing and the problem can be reduced to a quadrature,
\begin{equation}
\frac{1}{2} {y'}^2 + V(y) = E,
\label{eq:2.3}
\end{equation}
where $V(y) = \lambda \int_0^{y} f(s) \, ds$ and $E=V(y_m)$.

In \cite{BeDe98b}, Theorem 2.7, we proved the following variational characterization for the
principal eigenvalue of (\ref{eq:2.1}),(\ref{eq:2.2}).

\begin{theorem} Let the pair $(\lambda,y)$ be the principal solution (i.e., with $y(x) \ge 0$) of the two point boundary value problem
\begin{equation}
-\frac{d^2y}{dx^2} = \lambda f(y)
\label{eq:2.4}
\end{equation}
subject to $y'(0)=y(1)=0$. Let $y_m=y(0)$ be the sup-norm of the solution. Here, $f(y)$ is a general nonlinear term which is positive and continuous in $(0,y_m)$. 
Then,
\begin{equation}
\lambda[y_m]=\max_{g \in D} \frac{1}{2}
\frac{\left(\int_0^{y_m} g'(y)^{1/3} \, dy \right)^3} 
{\int_0^{y_m} f(y) g(y)\, dy},
\label{eq:2.5}
\end{equation}
where $D=\{ g \bigm| g \in C^1(0,y_m), g' > 0, g(0)=0\}$. Moreover, the maximum is attained at a unique (up to a multiplicative constant) $g \in D$. The maximizing $g$ satisfies
\begin{equation}
\frac{dg}{dy}= \frac{K}{(E-V(y))^{3/2}} \qquad \mbox{and $g(0)=0$.}
\label{eq:2.6}
\end{equation}
\end{theorem}

In what follows it is convenient to introduce
\begin{equation}
h(y) \equiv F(y_m) - F(y),
\label{eq:2.8}
\end{equation}
where
\begin{equation}
F(y) \equiv \int_0^y f(s) \, ds.
\label{eq:2.9}
\end{equation}

\begin{remark}
Using the optimal $g$, given through equation (\ref{eq:2.6}), in (\ref{eq:2.5}), integrating the 
denominator by parts, one gets the value of $\lambda[y_m]$ for the principal solution. This is given exactly by
\begin{equation}
\lambda[y_m]= \frac{1}{2}{\left(\int_0^{y_m} \frac{1}{\sqrt{h(s)}} \, ds \right)}^{2}.
\label{eq:2.7}
\end{equation}
After small manipulations, this expresion for $\lambda[y_m]$ is precisely the formula for the peiod given by (\ref{eq:1.1}) in the introduction.
\end{remark}

In the sequel, we will obtain a new variational characterization of $\lambda[y_m]$ which is the {\it dual} of (\ref{eq:2.5}). The dual principle is given by the following theorem.

\begin{theorem} Let the pair $(\lambda,y)$ be the principal solution (i.e., with $y(x) \ge 0$) of the two point boundary value problem (\ref{eq:2.1}), (\ref{eq:2.2}). 
Then,
\begin{equation}
\frac{1}{\lambda}= \max_{\sigma \in H}
\frac{2 \int_0^{y_m}(1- 2 \, \sigma(s)) \sqrt{(\sigma(s)+1)/h(s)} \, ds}
{\left(\int_0^{y_m} \sqrt{(\sigma(s)+1)/h(s)} \, ds \right)^3}.
\label{eq:2.10}
\end{equation}
where
$H=\{ \sigma \bigm| \sigma \in C(0,y_m), \sigma \ge 0\}$.
The maximum is attained at $\sigma \equiv 0$ and it is given precisely by 
\begin{equation}
\frac{1}{\lambda}= 2{\left(\int_0^{y_m} \frac{1}{\sqrt{h(s)}} \, ds \right)}^{-2}.
\label{eq:2.11}
\end{equation}
\end{theorem}
\begin{proof}
We will start from the original variational characterization of $\lambda$ embodied in Theorem 1,
and use the standard Fenchel--Moreau duality (see, e.g., the book by Ekeland and Temam \cite{EkTe99}) to obtain our dual principle. In order to simplify our calculations it is convenient to define $q(y)=g'(y)^{1/3}$, with $g \in D$. Integrating the denominator of (\ref{eq:2.5}) by parts, we obtain
\begin{equation}
\frac{1}{\lambda[y_m]}= 2 \min_{q} \frac{\int_0^{y_m} h(s) q^3 \, ds}
{\left( \int_0^{y_m}  q(s) \,ds \right)^3},
\label{eq:2.12}
\end{equation}
or, equivalently,
\begin{equation}
\frac{1}{\lambda[y_m]}= 2 \min_{q \in D'} \int_0^{y_m} h(s) q^3 \, ds,
\label{eq:2.12b}
\end{equation}
where $D'=\{ q \bigm| q \in C(0,y_m), q > 0, \int_0^{y_m} q(s) \, ds=1 \}$. 
Let $J[q] \equiv \int_0^{y_m} q(s)^3 h(s) \, ds$, and let $\hat J(t)$ be the Legendre transform of 
$J$, i.e.,
\begin{equation}
\hat J[t] = \sup_{q \in D'} \left( \int_0^{y_m} t(s) q(s) \, ds - J[q] \right).
\label{eq:2.13}
\end{equation}
In principle we should compute $\hat J(t)$ for all $t \in L^2(0,y_m)$. Doing so one finds that the function $t$ that maximizes $-\hat J(t)$ is $t\equiv 0$ (we leave the details to the reader). Having this in mind, and for the sake of simplicity, we may restrict to nonnegative functions $t(s)$, which we do in the sequel.
In order to simplify the computation of the Legendre transform of $J$ it is convenient to use the change of variables $t \to  \sigma$ given by $t = 3 \alpha \sigma$, with $\alpha$ given by
\begin{equation}
1=\sqrt{\alpha} \int_0^{y_m} \sqrt{\frac{1+\sigma(s)}{h(s)}} \, ds.
\label{eq:2.14}
\end{equation}
With this change of variables it follows from (\ref{eq:2.13}) and the definition of $J[q]$ 
that
\begin{equation}
\hat J [\sigma] = \alpha \sqrt{\alpha} \int_0^{y_m} \left( 2 \sigma -1 \right) \sqrt{\frac{1+\sigma(s)}{h(s)}} \, ds.
\label{eq:2.15}
\end{equation} 
Using (\ref{eq:2.14}) in (\ref{eq:2.15}) finally gives,
\begin{equation}
-\hat J[\sigma] = \frac
{\int_0^{y_m}(1- 2 \, \sigma(s)) \sqrt{(\sigma(s)+1)/h(s)} \, ds}
{\left(\int_0^{y_m} \sqrt{(\sigma(s)+1)/h(s)} \, ds \right)^3}.
\label{eq:2.16}
\end{equation}
Using Fenchel--Moreau duality, 
$$
\min_q J(q)= \max_{\sigma} \left[- \hat J(\sigma) \right],
$$ 
and the theorem follows from here.

\end{proof}

\section{An application: The Duffing equation}

To illustrate our variational principle and its dual, we will consider as an example the Duffing oscillator. Consider the two point boundary value problem
\begin{equation}
\ddot x= - \lambda(x + \delta x^3),
\label{eq:3.1}
\end{equation}
on the interval $(0,1)$ with $\dot x(0)=0$ and $x(1)=0$. Denote by $x_m=x(0)$ the sup--norm of the solution. We are only interested in the principal branch $(\lambda,x_m)$, i.e., in the positive solution. In terms of $\lambda$, the period of the oscillator is given by $4 \sqrt{\lambda}$. For this simple example one can compute the period of the oscillator in closed form in terms of complete elliptic integrals. However, since our purpose here is to illustrate the use of the dual variational principles, we will use simple trial functions in Theorems 1 and 2 above to compute upper and lower bounds on the period of the Duffing oscillator. There is a vast literature on the Duffing oscillator, and the use of perturbative schemes to obtain the period of the oscilator (see e.g., \cite{PeKlSc03} and the references therein). Most perturbative schemes yield divergent asymptotic series for the period which can, however, be converted into exponentially fast convergent series by Kleinert's variational perturbation theory \cite{Kl03}.

To begin with, we use Theorem 1 to get a lower bound on $\lambda$. 
As we pointed out in \cite{BeDe98b}, a good trial function is given by
\begin{equation}
g(x)=\frac{1}{x_m^2} \frac{x}{\sqrt{x_m^2-x^2}},
\label{eq:3.2}
\end{equation}
which is certainly in $D$. This trial function will yield a lower bound on $\lambda$ which 
gives an excellent agreement with the exact value of $\lambda$ near the bifurcation point $\lambda_1= \pi/2$. In fact, as it was illustrated in \cite{BeDe98b} it gives a good agreement for all values of the amplitude $x_m$. From (\ref{eq:3.1})  we get
\begin{equation}
g'(x)= \frac{1}{(x_m^2-x^2)^{3/2}}.
\label{eq:3.3}
\end{equation}
Using this trial function $g$ in Theorem 1, yields the lower bound
\begin{equation}
\lambda \ge \frac{\pi^2}{4} \frac{1}{1+ 3 \delta x_m^2 /4}.
\label{eq:3.4}
\end{equation}
If we denote by $\omega=2 \pi/T$, the above bound gives the following upper bound on $\omega$ for the Duffing oscillator,
\begin{equation}
\omega \le \sqrt{ 1 + \frac{3}{4} \delta x_m^2}.
\label{eq:3.5}
\end{equation}
We now proceed to compute upper bounds on $\lambda$ (i.e., lower bounds on $\omega$) using the dual principle given by Theorem 2. For the Duffing oscillator, $f(x)=x+\delta x^3$, and 
therefore $F(x)= x^2/2+ \delta x^4/4$. Hence, we can write
\begin{equation}
h(x) = h_0(1 + \delta z)
\label{eq:3.6}
\end{equation}
where
\begin{equation}
h_0=\frac{1}{2} (x_m^2 - x^2), \qquad \mbox{and} \qquad z=\frac{1}{2}(x_m^2+x^2).
\label{eq:3.7}
\end{equation}
In order to get simple, computable, bounds using the variational principle given by Theorem 2, it is convenient to express the trial function $\sigma$ in terms of a new variable, $p$, through the following relation,
\begin{equation}
\frac{\sigma + 1}{h} = \frac{1}{h_0} p^2.
\label{eq:3.8}
\end{equation}
Different approximations (i.e., different upper bounds on $\lambda$) are obtained by choosing 
different expressions for the trial function $p$. The simplest way to get computable bounds to different degree of accuracy is to use as trial $p$ the Taylor expansion of 
$(1+\delta z)^{-1/2}$ truncated to any even power. In the sequel we give the details for the 
first nontrivial truncation. Better approximations can be obtained in the same way. Choosing 
\begin{equation}
p(z)=1-\frac{1}{2} \delta z + \frac{3}{8} \delta^2 z^2,
\label{eq:3.9}
\end{equation}
(which corresponds to the Taylor expansion of $(1+\delta z)^{-1/2}$ truncated to second order),
yields
\begin{equation}
\sigma= \frac{(\delta z)^3}{64}(40 - 15 \delta z + 9 \delta^2 z^2),
\label{eq:3.10}
\end{equation}
which is nonnegative, and therefore belongs to $H$. Denoting by 
\begin{eqnarray}
A= & 67108864 -25165824 \delta x_m^2 + 14942208 {\delta}^2 x_m^4 - 41287680 \delta^3 x_m^6 \nonumber \\
& + 31073280 \delta^4 x_m^8 -24748416 \delta^5 x_m^{10} + 7192476 \delta^6 x_m^{12} -2202957 \delta^7 x_m^{14}, \nonumber
\end{eqnarray}
and
\begin{equation}
B=4 (256 - 96 \delta x_m^2 + 57 \delta^2 x_m^4)^3,
\nonumber
\end{equation}
then, 
\begin{equation}
\frac{1}{\lambda}  \ge \frac{4}{\pi^2} \frac{A}{B},
\label{eq:3.11}
\end{equation}
and 
\begin{equation}
\omega \ge \sqrt{\frac{A}{B}}.
\label{eq:3.12}
\end{equation}

\begin{figure}[h]
\begin{center}
\epsfig{file=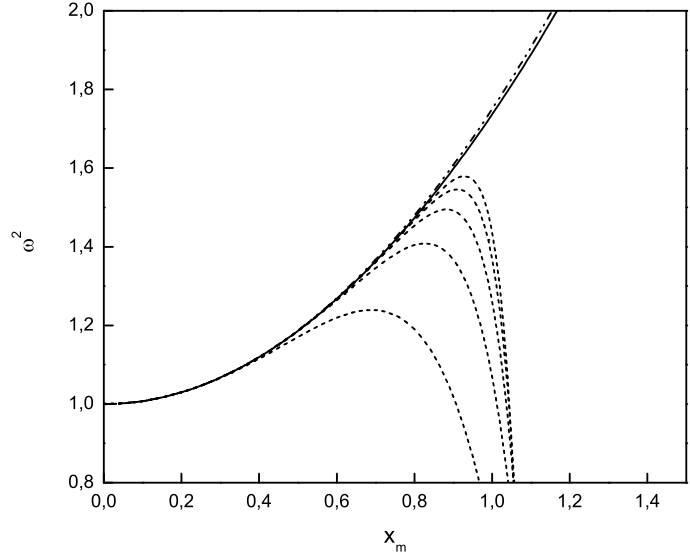,height=9.0truecm}
\caption{The solid line is the exact solution. The dot--dot--dash line is the upper bound, whereas the dashed lines are lower bounds from the variational principles.}
\end{center}
\end{figure}

In Figure 1, we plot the exact value (solid line) of $\omega\equiv 2 \pi/T$ for the Duffing oscilator as well 
as the upper bound (\ref{eq:3.5}) and the lower bound (\ref{eq:3.12}). The upper bound is 
an excellent approximation for $\omega$ for all amplitudes, whereas more complicated trial functions have to be used to get better lower bounds on $\omega$. In the figure we also
include the lower bounds obtained by truncating $(1+\delta z)^{-1/2}$ to degree $4$, $6$, $8$, 
and $10$ as a function of $z$. It is interesting to point out that all these lower bounds
are precisely the curves for $\omega$ obtained by standard perturbation theory.

{}

\end{document}